\documentclass[final,5p,times,twocolumn]{elsarticle} 
\DeclareMathAlphabet{\mathpzc}{OT1}{pzc}{m}{it}
\usepackage{lineno}
\usepackage{setspace}
\usepackage{graphicx}
\usepackage{amssymb}
\usepackage{amsmath}
\usepackage{amsthm}
\usepackage[colorlinks=true]{hyperref}
\usepackage[all]{hypcap}
\usepackage{mathrsfs}

\usepackage[percent]{overpic}
\usepackage{fixltx2e}
\usepackage{epsfig}
\usepackage{verbatim}
\usepackage{multirow}
\usepackage{float}
\usepackage{array}
\usepackage{alltt}
\usepackage{relsize}
\usepackage{upgreek}

\newcommand{\D}{\mathrm{d}}
\newcommand{\diag}{\mathrm{diag}}
\newcommand{\mn}{\mathrm{min}}
\newcommand{\mx}{\mathrm{max}}

\newcommand{\en}{E}
\newcommand{\etof}{\mathcal{E}}
\newcommand{\x}{X}
\newcommand{\q}{Q}
\newcommand{\tf}{t}

\newcommand{\rmx}{\boldsymbol{\mathrm{R}}}
\newcommand{\pmx}{\boldsymbol{\mathrm{P}}}
\newcommand{\ppmx}{\boldsymbol{\mathcal{P}}}
\newcommand{\dppmx}{\boldsymbol{\delta}\ppmx}
\newcommand{\emx}{\boldsymbol{\mathrm{\en}}}
\newcommand{\xmx}{\boldsymbol{\mathrm{\x}}}

\newcommand{\mmx}{\boldsymbol{\mathrm{M}}}
\newcommand{\amx}{\boldsymbol{\mathrm{A}}}
\newcommand{\bmx}{\boldsymbol{\mathrm{B}}}
\newcommand{\lmx}{\boldsymbol{\mathrm{L}}}
\newcommand{\qmx}{\boldsymbol{\mathrm{\q}}}

\journal{Nuclear Instruments and Methods A}
\begin{document}
\begin{frontmatter}

\title{A direct method for unfolding the resolution function from measurements of neutron induced reactions}


\author[a]{P.~\v{Z}ugec\corref{cor1}}\ead{pzugec@phy.hr}
\author[b]{N.~Colonna}
\author[c,d]{M.~Sabate-Gilarte}
\author[c]{V.~Vlachoudis}
\author[e,f]{C.~Massimi}
\author[d]{J.~Lerendegui-Marco}
\author[g]{A.~Stamatopoulos}
\author[c,h,i]{M.~Bacak}
\author[c]{S.G.~Warren}

\address[a]{Department of Physics, Faculty of Science, University of Zagreb, Croatia}
\address[b]{Istituto Nazionale di Fisica Nucleare, Sezione di Bari, Italy}
\address[c]{European Organization for Nuclear Research (CERN), Geneva, Switzerland}
\address[d]{Universidad de Sevilla, Spain}
\address[e]{Istituto Nazionale di Fisica Nucleare, Sezione di Bologna, Italy}
\address[f]{Dipartimento di Fisica e Astronomia, Universit\`{a} di Bologna, Italy}
\address[g]{National Technical University of Athens (NTUA), Greece}
\address[h]{Atominstitut der \"{O}sterreichischen Universit\"{a}ten, Technische Universit\"{a}t Wien, Austria}
\address[i]{Irfu, CEA, Universit\'{e} Paris-Saclay, Gif-sur-Yvette, France}

\author{\linebreak The n\_TOF Collaboration\fnref{fn1}} 
\cortext[cor1]{Corresponding author. Tel.: +385 1 4605552}
\fntext[fn1]{www.cern.ch/ntof}

\begin{abstract}
The paper explores the numerical stability and the computational efficiency of a direct method for unfolding the resolution function from the measurements of the neutron induced reactions. A detailed resolution function formalism is laid out, followed by an overview of challenges present in a practical implementation of the method. A special matrix storage scheme is developed in order to facilitate both the memory management of the resolution function matrix, and to increase the computational efficiency of the matrix multiplication and decomposition procedures. Due to its admirable computational properties, a Cholesky decomposition is at the heart of the unfolding procedure. With the smallest but necessary modification of the matrix to be decomposed, the method is successfully applied to system of $10^5\times10^5$. However, the amplification of the uncertainties during the direct inversion procedures limits the applicability of the method to high-precision measurements of neutron induced reactions.
\end{abstract}

\end{frontmatter}

\section{Introduction}
\label{sec:chap1}

The resolution function of the neutron beam is an inherent characteristic of the neutron production facilities, such as the neutron time of flight facility n\_TOF \cite{ntof} at CERN. In particular, the resolution function $R(\tf,\en)$ is the distribution of the neutron flight time $\tf$ (over the flight path of a fixed length $L$) for neutrons of a given kinetic energy $\en$. As opposed to the idealized one-to-one correspondence between time of flight and neutron energy, the flight time can vary due to several experimental effects: (1)~the time spread of the primary beam of charged particles producing the neutron beam; (2)~the variable moderation time in the target-moderator assembly; (3)~the geometry of the neutron propagation along the beam line of finite length and diameter. In case of the n\_TOF facility, the width of the primary 20~GeV proton beam from the CERN Proton Synchrotron is 7~ns RMS. The target-moderator assembly consists of a massive lead spallation target, 40~cm in length and 60~cm in diameter, surrounded by a cooling system comprised of 1~cm layer of demineralized water and additional 4~cm of either borated or demineralized water. The n\_TOF facility features two experiential areas: Experimental Area 1 (EAR1 \cite{ntof}) at the horizontal distance of 185~m from the spallation target, and Experimental Area 2 (EAR2 \cite{ear2_1,ear2_2,ear2_3}) at the vertical distance of 20~m from the same target. Therefore, two separate resolution functions must be taken into account, one for each experimental area.

At n\_TOF the resolution function of the neutron beam has been given consideration ever since the very conception of the facility \cite{ntof0}, throughout the start of its operation \cite{rf00,rf0}, to the present day \cite{ntof,ear2_1,rf1,rf2}. Now, after entering n\_TOF-Phase3 -- the third phase in the operation of the n\_TOF facility, marked by the successful completion, commissioning and the start of the operation of EAR2 -- the resolution function considerations are being pursued with greater fervour than ever before \cite{ear2_1,rf1,rf2}. These efforts are motivated by the requirement for the optimum quality in processing the high precision experimental data obtained at n\_TOF, which are regularly a key input to the development of nuclear technologies and are of importance to the general scientific community \cite{nuc1,nuc2}.

The reliable evaluation of the resolution function throughout wide neutron energy range may only be obtained through the dedicated simulations of the neutron production and their propagation through the target-moderator assembly. However, the only way of evaluating the reliability of the simulated results is by benchmarking them against the experimental data, by means of applying the simulated resolution function to the estimated reaction yield of well known resonances (such as the neutron capture resonances of $^{25}$Mg, $^{56}$Fe, $^{197}$Au) and comparing them with the available measurements \cite{rf1,rf2}. Once the reliability of the resolution function has been confirmed, it may be used either in its numerical format, or it may be fitted to an appropriate analytical form. The form from Ref.~\cite{ntof} is used at n\_TOF, while Ref.~\cite{sammy} lists some other forms widely used at other neutron production facilities.

Ultimately, the resolution function in either form needs to be applied to the experimental data, in order to decouple its effect from the measurements, in particular from the measured resonances in neutron induced reactions. One way of performing this task is by relying on specialized codes like SAMMY \cite{sammy} or REFIT \cite{refit}, which use the R-matrix formalism to fit the experimental data to the parameterized form of the underlying cross section. In the process the parameterized cross section is folded or convoluted with the self-shielding, multiple scattering, Doppler broadening and, ultimately, the resolution function effects in order to reproduce the observable and, indeed, measured reaction yield. This is certainly a very robust, stable and reliable method. Notwithstanding, in this work we explore an alternative approach, its applicability and limitations. While the SAMMY/REFIT approach is evidently a forward application of the resolution function to the assumed pure form of the underlying cross section, we seek to address the inverse problem: how to directly unfold the resolution function from the measurements, starting from the data already affected by it.

Section~\ref{theory} presents the theoretical introduction to the resolution function, establishing the central problem to be solved. Section~\ref{unfolding} reports on the successfully applied unfolding procedure. Section~\ref{uncertainty} addresses the propagation of uncertainties, which is a crucial issue for any of the direct unfolding procedures. Section~\ref{conclusion} sums up the conclusions of this work. \ref{matrix} presents a matrix storage scheme central to this work and lays out the important technical procedures to be performed prior to the unfolding itself. In addition, \ref{multiple} addresses the applicability of the unfolding procedure in the presence of pronounced multiple scattering effects.

\section{Theoretical framework}
\label{theory}

\subsection{Resolution function}

Let us consider the resolution function $R_\x(\x,\en)$ expressed as a function of some general kinematic parameter $\x$. For the moment we identify $\x$ with the neutron time of flight: $\x=\tf$. Then the precise definition of the resolution function takes the form:
\begin{linenomath}\begin{equation}
\label{rf}
R_\x(\x,\en)\equiv\frac{\D P_\x(\x,\en)}{\D \x}
\end{equation}\end{linenomath}
where $\D P_\x(\x,\en)$ is the probability\footnote{The resolution function is normalized over $\x$ for every value of $\en$:
\begin{linenomath}\begin{equation*}
\int_{-\infty}^\infty R_\x(\x,\en)\D \x=1\quad\forall \en
\end{equation*}\end{linenomath}
In principle, the lower limit of integration may not be 0, depending on the definition of the kinematic parameter $\x$. For example, negative values of \mbox{$\x=\tf$} are introduced by an arbitrary selection of the nominal initial moment $\tf=0$ for the neutron production. If the initial moment is selected such that no neutrons are produced prior to $\tf=0$, only then may the lower limit of integration be set to 0.} for the neutron of kinetic energy $\en$ to have the corresponding kinematic parameter $\x$ within the interval $\D \x$, i.e. to have the time of flight $\tf$ within the time of flight interval $\D \tf$. Aside from the time of flight, one may express the resolution function in terms of the reconstructed energy $\etof$, which is the equivalent neutron kinetic energy calculated from the relativistic time-energy relation:
\begin{linenomath}\begin{equation}
\label{etof}
\etof=m c^2\left\{\left[1-\left(\frac{L}{c\tf}\right)^2\right]^{-1/2}-1\right\}
\end{equation}\end{linenomath}
with $m$ as the neutron mass, $c$ as the speed of light in vacuum and $L$ as the mean neutron flight path. Another commonly used parameterization is the one defining the effective neutron flight path $\lambda$:
\begin{linenomath}\begin{equation}
\label{lambda}
\lambda\equiv v_\en \tf-\lambda_0=\frac{\sqrt{\left(\en/m c^2\right)^2+2\en/m c^2}}{1+\en/m c^2}\times c\tf-\lambda_0
\end{equation}\end{linenomath}
where $v_\en$ is the neutron speed calculated from a true neutron kinetic energy $\en$. In principle, the value of the shift parameter $\lambda_0$ is arbitrary, but two conventions are commonly used: $\lambda_0=0$ or $\lambda_0=L$. Under the assumption of $\lambda_0=0$, the effective path length $\lambda$ corresponds to the entire path length that the neutron of energy $\en$ would need to traverse to arrive at the measurement position at time $\tf$. On the other hand, when $\lambda_0=L$ is assumed, then $\lambda$ is interpreted as an effective path the neutron needs to traverse within the target-moderator assembly. Whichever kinematic parameter $\x\in\{\tf,\etof,\lambda\}$ is selected for expressing the resolution function, the probability conservations requires the following to hold:
\begin{linenomath}\begin{equation}
R_\tf(\tf,\en)\;|\D \tf|=R_{\etof}(\etof,\en)\;|\D \etof|=R_\lambda(\lambda,\en)\;|\D \lambda|
\end{equation}\end{linenomath}
allowing for a simple transformation between different forms of the resolution function.

A note should be taken of the relation between the resolution functions at different paths, in particular at the nominal neutron production point $L=0$ and at some specific distance $L>0$ corresponding to a measurement position. One may be tempted to use the function $R_\tf^{(L=0)}(\delta \tf,\en)$ (with $\delta \tf$ as effective production time relative to the nominal initial moment $\tf=0$) in order to reconstruct the resolution function $R_\tf^{(L>0)}(\tf,\en)$ at any flight path $L$ simply by propagating the neutron in time by the time of flight $\tf_\en=L/v_\en$, with $v_\en$ as the neutron speed defined by Eq.~(\ref{lambda}). However, the inequality:
\begin{linenomath}\begin{equation}
R_\tf^{(L>0)}(\tf,\en) \neq R_\tf^{(L=0)}(\tf-\tf_\en,\en)
\end{equation}\end{linenomath}
holds in general, for two reasons. One is that many neutrons reach the measurement position propagating under a slight angle $\theta$ relative to the shortest flight path, making their actual path length slightly longer ($L/\cos\theta$). The other reason is that the neutrons exiting the primary neutron source -- directly giving rise to $R_\tf^{(L=0)}(\delta \tf,\en)$ -- may interact with any material along the flight path, thus being lost, delayed and/or producing secondary neutrons with kinematic parameters uncorrelated to their own initial ones. Therefore, the final resolution function $R_\tf^{(L>0)}(\tf,\en)$ may have additional contributions, not inherited from $R_\tf^{(L=0)}(\delta \tf,\en)$ at the neutron production point. For this reason, throughout this work we will only consider the resolution functions at the measurement position.

\subsection{Continuous representation}


Let us first lay down the formalism of continuous resolution functions, establishing the connection between the measured spectrum $S_\x(\x)$ expressed as a function of an experimentally accessible kinematic parameter $\x$, and the underlying spectrum $S_\en(\en)$ dependent on the true neutron kinetic energy $\en$. The considerations are equally valid whether the selected kinematic parameter is the neutron time of flight ($\x=\tf$) or the reconstructed neutron energy ($\x=\etof$).

Let $\D^2 N(\x,\en)$ be the number of reactions caused by the neutrons of energy $\en$ and detected with the kinematic parameter $\x$. We define the corresponding spectrum $\mathcal{S}(\x,\en)$ of detected counts as:
\begin{linenomath}\begin{equation}
\label{spec}
\mathcal{S}(\x,\en)\equiv \frac{\D^2 N(\x,\en)}{\D\x \D\en}
\end{equation}\end{linenomath}
We wish to reconstruct the spectrum $S_\en(\en)$, unaffected by the resolution function:
\begin{linenomath}\begin{equation}
S_\en(\en)\equiv\frac{\D N_\en(\en)}{\D\en}=\int_{-\infty}^\infty \mathcal{S}(\x,\en)\D\x
\end{equation}\end{linenomath}
starting from the measured spectrum $S_\x(\x)$:
\begin{linenomath}\begin{equation}
\label{sx}
S_\x(\x)\equiv\frac{\D N_\x(\x)}{\D \x}=\int_0^\infty \mathcal{S}(\x,\en)\D\en
\end{equation}\end{linenomath}
Since $\D N_\en(\en)=S_\en(\en)\D\en$ is the total number of detected counts caused by the neutrons of energy $\en$, the number of detections with a particular value of the kinematic parameter $\x$ equals:
\begin{linenomath}\begin{equation}
\label{eq_d2}
\D^2 N(\x,\en)=\D N_\en(\en)\times \D P_\x(\x,\en)=S_\en(\en)R_\x(\x,\en)\D\x \D\en
\end{equation}\end{linenomath}
where we have taken advantage of the probability $\D P_\x(\x,\en)$ and the definition of the resolution function $R_\x(\x,\en)$ from Eq.~(\ref{rf}). Inserting this expression into Eq.~(\ref{spec}) yields:
\begin{linenomath}\begin{equation}
\mathcal{S}(\x,\en)=S_\en(\en)R_\x(\x,\en)
\end{equation}\end{linenomath}
which, in turn, via Eq.~(\ref{sx}) leads to:
\begin{linenomath}\begin{equation}
\label{integral}
S_\x(\x)=\int_0^\infty S_\en(\en)R_\x(\x,\en)\D\en
\end{equation}\end{linenomath}
otherwise known as a Fredholm integral equation of the first kind.

\subsection{Discretized representation}

In practical data analysis, the amount of data is finite and a discretized version of Eq.~(\ref{integral}) is required:
\begin{linenomath}\begin{equation}
\label{discrete}
S_\x(\x_i)=\sum_j S_\en(\en_j) R_\x(\x_i,\en_j)\Delta \en_j
\end{equation}\end{linenomath}
with indices $i$ and $j$ denoting the corresponding histogram bins. In this case the discretized version of the resolution function is simply the average over both the source and the destination bin:
\begin{linenomath}\begin{equation}
\label{new_def}
R_\x^{(\Delta>0)}(\x_i,\en_j)=\frac{1}{\Delta \x_i \Delta\en_j}\int_{\Delta \x_i}\int_{\Delta\en_j}R_\x^{(\Delta\rightarrow0)}(\x,\en) \D\x\D\en
\end{equation}\end{linenomath}
with $\Delta \x_i$ and $\Delta \en_j$ as the corresponding bin widths. $R_\x^{(\Delta>0)}$ on the left side of Eq.~(\ref{new_def}) denotes the discretized representation, while $R_\x^{(\Delta\rightarrow0)}$ from the right side the continuous one. By constructing the vectors $\vec{S}_\en$ and $\vec{S}_\x$ of the histogrammed spectra such that $[\vec{S}_\en]_i\equiv S_\en(\en_i)$ and $[\vec{S}_\x]_i\equiv S_\x(\x_i)$, Eq.~(\ref{discrete}) may be rewritten in a compact matrix form:
\begin{linenomath}\begin{equation}
\label{submaster}
\vec{S}_\x=\rmx \:\vec{S}_\en
\end{equation}\end{linenomath}
with the resolution function matrix $\rmx$ defined as:
\begin{linenomath}\begin{equation}
\label{rdef}
\rmx_{ij}\equiv R_\x(\x_i,\en_j)\Delta \en_j =\frac{\Delta\en_j}{\Delta \x_i}\Delta P_\x(\x_i,\en_j)
\end{equation}\end{linenomath}
Evidently, $\Delta P_\x$ is the discretized version of the probability $\D P_\x$ from Eq.~(\ref{rf}).

Our ultimate goal -- obtaining the spectrum $S_\en(\en)$ from the measured spectrum $S_\x(\x)$ -- is now reduced to finding a solution of the linear system from Eq.~(\ref{submaster}). At this point it is useful to note that the matrix $\rmx$ may be rewritten as:
\begin{linenomath}\begin{equation}
\label{decomp}
\rmx=\xmx^{-1}\pmx\emx
\end{equation}\end{linenomath}
where $\xmx\equiv\diag[\Delta \x_i]$ and $\emx\equiv\diag[\Delta \en_i]$ are diagonal matrices and the probability matrix $\pmx$ is simply:
\begin{linenomath}\begin{equation}
\pmx_{ij}\equiv \Delta P_\x(\x_i,\en_j)=\frac{1}{\Delta \en_j}\int_{\Delta \en_j}\D \en\int_{\Delta \x_j} \D P_\x(\x,\en)
\end{equation}\end{linenomath}
The task of inverting $\rmx$ from Eq.~(\ref{decomp}): $\rmx^{-1}=\emx^{-1}\pmx^{-1}\xmx$ is thus translated into that of inverting $\pmx$. The advantage of this approach is the unification of the numerical stability issues, regardless of the selection of the kinematic parameter $X$. If the reconstructed neutron energy ($\x=\etof$) is selected, then the relevant values of the fractional term from Eq.~(\ref{rdef}) are close to unity ($\Delta \en_j/\Delta\etof_i\approx1$), making the average magnitude of $\rmx_{ij}$ values basically uniform over the entire range of $\en$. On the other hand, when the neutron time of flight ($\x=\tf$) is selected, there is a massive variation in magnitude if the data span the wide range in $E$, which is easily shown by taking advantage of the time-energy relation from Eq.~(\ref{etof}):
\begin{linenomath}\begin{equation}
\frac{\Delta\en_j}{\Delta \tf_i}\approx \left|\frac{\D \tf}{\D \etof}\right|_{\etof=\en_j}^{-1}=\frac{{\sqrt{\en_j(2+\en_j/mc^2)}}^{\:3}}{L\sqrt{m}} \xrightarrow{\en_j\ll mc^2} \frac{1}{L}\sqrt{\frac{8\en_j^3}{m}}
\end{equation}\end{linenomath}
Adopting $\pmx$ for the numerical treatment circumvents this inconsistency, since $\pmx_{ij}$ are the true probabilities, subject to the normalization $\sum_{i}\pmx_{ij}=1$ for every $j$.

Having selected $\pmx$ in place of the initial $\rmx$, we may note that by inserting Eq.~(\ref{decomp}) into Eq.~(\ref{submaster}) one can make a transition from differential spectra $\vec{S}_\en$ and $\vec{S}_\x$ directly into the spectra of counts:
\begin{linenomath}\begin{align}
\begin{split}
\vec{N}_\en&\equiv \emx\:\vec{S}_\en\\
\vec{N}_\x&\equiv \xmx\:\vec{S}_\x
\end{split}
\end{align}\end{linenomath}
Evidently, the content of these vectors is just the amount of counts from each corresponding histogram bin: \mbox{$[\vec{N}_\en]_i=\Delta N_\en(\en_i)$} and \mbox{$[\vec{N}_\x]_i=\Delta N_\x(\x_i)$}. Thus the central equation of interest becomes:
\begin{linenomath}\begin{equation}
\label{master}
\vec{N}_\x=\pmx\:\vec{N}_\en
\end{equation}\end{linenomath}
whose solution $\vec{N}_\en$ we seek to compute and analyze.

Throughout the following sections we deal with many well established concepts in solving the linear system of equations. While the literature on the subject is vast, we point the reader to Refs.~\cite{matrix,numc} for a highly comprehensive and practical overview of the subject.

\section{Unfolding procedure}
\label{unfolding}

The potential challenges in solving Eq.~(\ref{master}) arise when we consider the size of the system we aim to handle. The neutron capture measurements from EAR1 often span the energy range from $\sim$10~meV up to $\sim$1~MeV, i.e. covering 8 orders of magnitude in energy. Considering that the binning used for these data often reaches 5000 bins per decade \cite{fede}, we need to solve systems of the size $(4\times10^4)\times(4\times10^4)$, at the very least. Furthermore, the method should be of sufficient generality to make the transition between the spectra $\vec{N}_\en$ and $\vec{N}_\x$ not only of differently distributed bins, but also of the different number of bins, implying that $\pmx$ should not necessarily be a square matrix. In this case the so-called pseudoinverse of $\pmx$ may still be constructed and expressed as: $\pmx^{-1}=(\pmx^\top\pmx)^{-1}\pmx^\top$, with $(\cdot)^\top$ standing for the matrix transposition. When applied, the pseudoinverse extracts the best solution to the system of equations, in a least-squares sense. However, this operation is only meaningful if the number of elements $n_\en$ in the solution vector $\vec{N}_\en$ is not greater than the number of elements $n_\x$ in $\vec{N}_\x$. In other words, if $\pmx$ is of the size $n_\x \times n_\en$, then $n_\x\ge n_\en$ should be satisfied in order for a unique solution $\vec{N}_\en$ to exist \cite{numc}.

\begin{figure}[b!]
\includegraphics[width=1.0\linewidth]{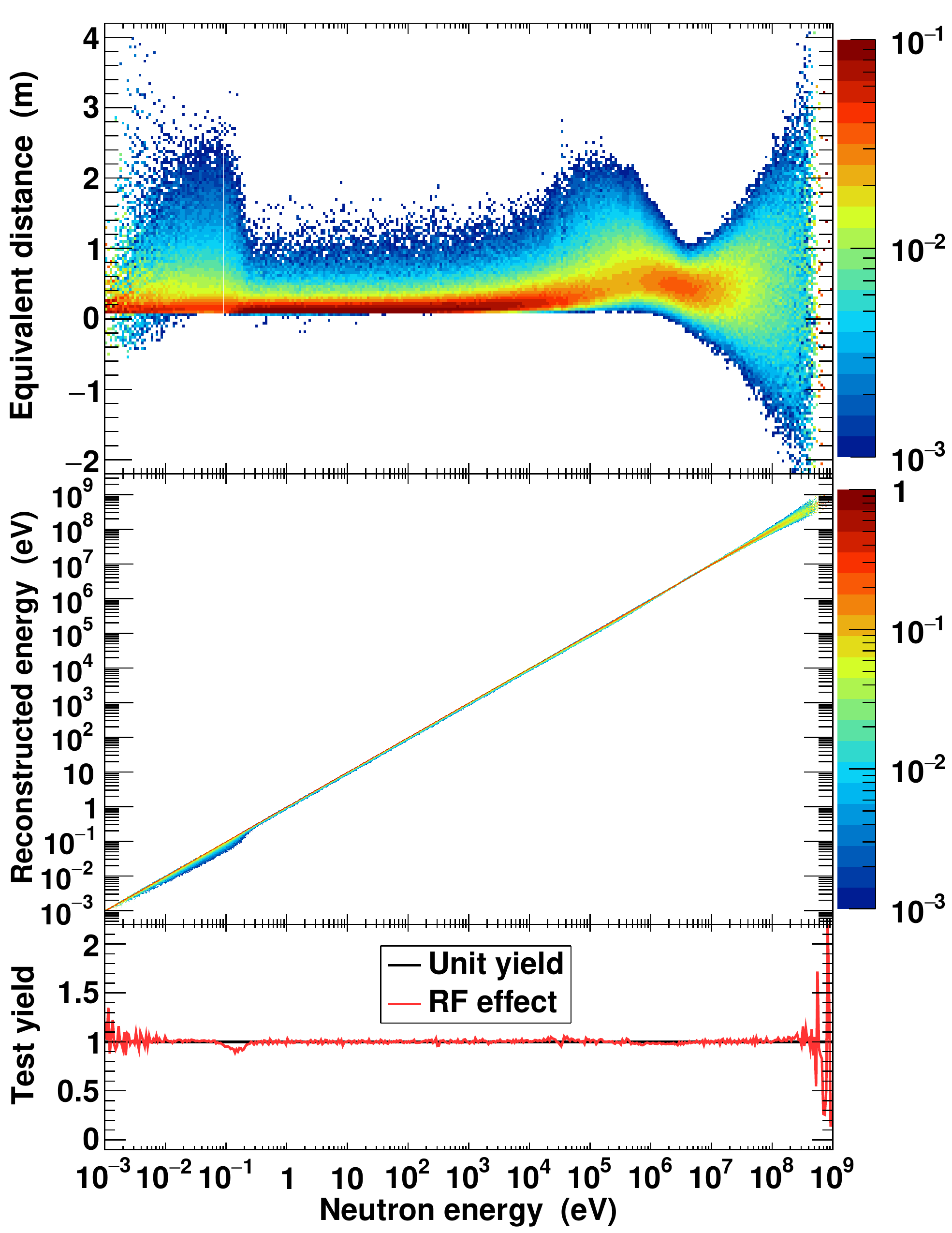}
\caption{Probability matrix $\pmx$, representative of the resolution function for the EAR2 of the n\_TOF facility. The numerical matrix has been adopted directly from the raw Geant4 simulations. Top panel: $\pmx$ as a function of an effective neutron flight inside the target-moderator assembly (equivalent distance, for short). Middle panel: $\pmx$ as function of the neutron energy reconstructed from the neutron time of flight. Bottom panel: the unit-spectrum folded by the adopted probability matrix.}
\label{fig1}
\end{figure}

Figure~\ref{fig1} shows the resolution function for the EAR2 of the n\_TOF facility, obtained by Geant4 simulations \cite{rf2}, which we have selected for presentation throughout this work. The top panel shows the probability matrix $\pmx$ as a function of an effective flight path length $\lambda$, which may be easily translated into the time of flight $\tf$ by Eq.~(\ref{lambda}), or further into the reconstructed neutron energy $\etof$ by Eq.~(\ref{etof}). In presenting the graphical examples from this work we show the reconstructed neutron energy, as this enables the direct comparison of the folded spectra with the ones from before and after unfolding. Thus, the middle panel from Fig.~\ref{fig1} shows the same probability matrix $\pmx$ as a function of $\etof$. Alternatively, Fig.~13 from Ref.~\cite{ear2_1} presents an example of the (unnormalized) $\pmx$ as a function of the time of flight~$\tf$ (obtained, in that work, by FLUKA simulations). Finally, the bottom plot from Fig.~\ref{fig1} shows the effect of the adopted probability matrix when applied to the unit-spectrum. The rough shape of the folded spectrum reflects the fact that we have adopted the probability matrix directly from raw simulations, without much refinements. This serves our goals well, as we wish to investigate the numerical stability of the unfolding procedure even in case of a challenging $\pmx$, which does not necessarily yield a smooth and "well-behaved" spectrum for unfolding.

Immediately evident from the middle panel of Fig.~\ref{fig1} (as well as Fig.~13 from Ref.~\cite{ear2_1}) is the banded structure of $\pmx$ -- or, equivalently, of the resolution function matrix $\rmx$ -- which we will take advantage of in order to increase the efficiency of the matrix computations and to reduce the matrix storage requirements. To this end we adopt the storage scheme from \ref{storage}, which we apply to all matrices referred to in this work. In addition, several important technical procedures need to be performed before solving the system of Eq.~(\ref{master}), all of which are described in \ref{preparation}.

\subsection{Solution to the system of equations}
\label{cholesky}

The most reliable method of solving the system from Eq.~(\ref{master}), in the sense of numerical stability and a prospect of success, is the Singular Value Decomposition (SVD) \cite{matrix,numc} of $\pmx$. Moreover, SVD is directly applicable even when $\pmx$ is not square, which is the level of generality that we wish to maintain. However, for the systems of the size that we aim to handle (of the order $10^4\times10^4$ or even $10^5\times10^5$) the SVD methods become prohibitively expensive to execute. Furthermore, while SVD may be used to directly construct the inverse $\pmx^{-1}$, for large systems this approach becomes of limited use. The main reason is that the inverse of banded matrices -- bandedness being the feature of $\pmx$ that we heavily rely on to be able to store it efficiently -- is dense in general, i.e. does not possess or even resemble the banded structure of the original matrix. Thus, significantly higher, even prohibitive memory resources would be needed just to store the inverse. Fortunately, there is a variety of alternative methods available for obtaining the solution of Eq.~(\ref{master}). Some entirely avoid the construction of any intermediate matrix, and some -- if successful -- construct intermediate matrices of favorable properties (preserving, for example, the banded structure), while utilizing far simpler computational schemes than SVD. The methods of the former class -- requiring only the original matrix $\pmx$ -- are usually iterative in nature and we reflect on them in Section~\ref{iterative}. In this section, however, we focus on a method of the latter class, namely the Cholesky decomposition of a symmetric, positive definite matrix.

The system from Eq.~(\ref{master}) can always be reduced to a so-called system of normal equations by applying the transpose $\pmx^\top$, so that: $\pmx^\top\vec{N}_\x=\pmx^\top\pmx\:\vec{N}_\en$. Introducing the notation $\ppmx$ for a new core-matrix, and $\vec{\mathcal{N}}_\x$ for a modified spectrum of measured counts, the system to be solved takes the form:
\begin{linenomath}\begin{equation}
\label{new_master}
\left.\begin{array}{c}\ppmx\equiv \pmx^\top\pmx\\\vec{\mathcal{N}}_\x\equiv\pmx^\top\vec{N}_\x\end{array}\right\} \quad\Rightarrow\quad \vec{\mathcal{N}}_\x=\ppmx \vec{N}_\en
\end{equation}\end{linenomath}
The new core-matrix $\ppmx$ is not only square, but also symmetric and positive definite, making it a prime candidate for a Cholesky decomposition. In addition, for banded $\pmx$, $\ppmx$ will also be banded, though with a wider bandwidth. One issue to take note of is that it is not usually recommended to substitute the system from Eq.~(\ref{master}) by the one from Eq.~(\ref{new_master}), as the matrix $\ppmx$ has a greater condition number than $\pmx$, bringing it closer to being ill-conditioned, thus making the decomposition and inversion procedures more susceptible to failure \cite{matrix,numc}. However, in light of our requirements (non-square matrix, the size of the system), not many alternatives remain available. Furthermore, the benefits of the Cholesky decomposition -- if it can be successfully executed -- are so overwhelming that any attempt at it is certainly justified. These benefits include: (1)~extremely simple and efficient algorithm for performing the decomposition itself; (2)~extremely efficient procedure for obtaining the solution to the linear system of equations from a decomposed matrix, by means of a forward and backward substitution; (3)~the Cholesky decomposition inherits the banded structure of the decomposed matrix \cite{matrix}, not only further reducing the number of required operations, but also allowing for a compact storage of the decomposition result. If the decomposition is of sufficient numerical accuracy, the direct solution to Eq.~(\ref{new_master}) may be kept as the final result, or at least as a good initial guess for the iterative methods (Section~\ref{iterative}), accelerating their convergence towards the true solution.

The Cholesky decomposition itself is of the form:
\begin{linenomath}\begin{equation}
\ppmx=\lmx\lmx^\top
\end{equation}\end{linenomath}
with $\lmx$ as the lower triangular matrix. Employing the default algorithm \cite{matrix,numc}, we indeed observe the failure of the decomposition when $\ppmx$ reaches the size of approximately $10^3\times10^3$. The failure bears an unmistakable signature, as it is realized through the appearance of negative terms under the square roots involved in the calculation. However, even if the solution $\vec{N}_\en$ to Eq.~(\ref{new_master}) is successfully found by means of the Cholesky decomposition, in the presence of the roundoff errors caused by the finite-precision arithmetic this solution does not solve exactly the starting  Eq.~(\ref{new_master}), but rather the system $(\ppmx+\dppmx)\vec{N}_\en$ where $\dppmx$ is a perturbation to $\ppmx$ \cite{wilkinson,memoriam}. This implies that if the decomposition fails for $\ppmx$, some perturbation $\dppmx$ could be manually added to it in order to promote it towards the closest machine-representable matrix for which the decomposition still succeeds. This matrix upgrade may go even further, to the point where the decomposition not only reconstructs $\ppmx$ as well as possible, but also yields the most accurate inverse $\ppmx^{-1}$. In an attempt to promote the positive definiteness, we amplify the diagonal elements of $\ppmx$ by a small adjustable factor~$\epsilon$, meaning that $\dppmx=\diag[\epsilon\,\ppmx_{ii}]$. This modification may be directly implemented into the equations for calculating the elements of the Cholesky factor $\lmx$:
\begin{linenomath}\begin{align}
\begin{split}
\lmx_{ii}&=\sqrt{\ppmx_{ii}(1+\epsilon)-\sum_{k=0}^{i-1}\lmx_{ik}^2}\\
\lmx_{ij}&=\frac{1}{\lmx_{ii}}\left(\ppmx_{ij}-\sum_{k=0}^{i-1}\lmx_{ik}\lmx_{jk}\right)
\end{split}
\end{align}\end{linenomath}
Indeed, this practice stabilizes the decomposition of $\ppmx$ even for the values of $\epsilon$ close to the limit of the numerical precision (\mbox{$\epsilon\approx10^{-6}$} for type \textit{float} and \mbox{$\epsilon\approx10^{-15}$} for type \textit{double} from C++). However, for the matrices of the size up to $10^5\times10^5$ (and for our particular resolution function) we found the optimal value of $\epsilon=10^{-4}$. In principle, this factor depends on the particular binning of the resolution function matrix. But if a unique value is sought, the optimization is best performed using a matrix as large as possible, since the smaller matrices not only suffer less from the stability issues but are also less sensitive to the value of $\epsilon$. As suggested earlier, the method of optimization relies on observing the quality of the solution to Eq.~(\ref{new_master}), which visibly degrades for values of $\epsilon$ sufficiently far from the optimal one.

\begin{figure}[t!]
\includegraphics[width=1.0\linewidth]{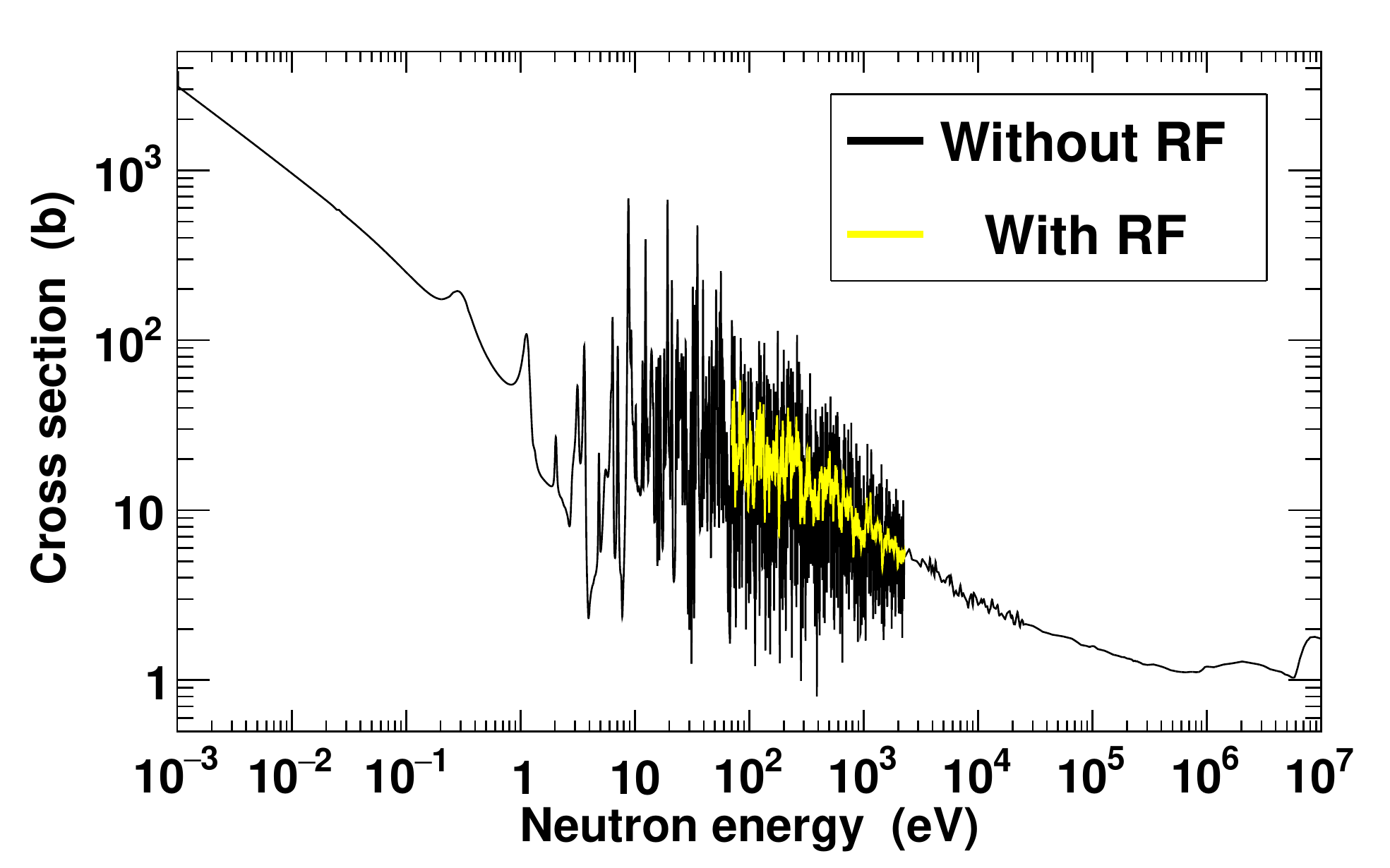}
\caption{Cross section for the $^{235}$U($n$,f) reaction, from the ENDF/B-VII.1 database. For visual purposes, the smoothing effect of the resolution function for the EAR2 of the n\_TOF facility is displayed only within the dense resolved resonance region.}
\label{fig2}
\end{figure}

Figure~\ref{fig2} shows the cross section of the $^{235}$U($n$,f) reaction, adopted from the ENDF/B-VII.1 database \cite{endf}, that we have selected for presenting the results of the unfolding procedure. This cross section features all the common characteristics of a neutron reaction cross section -- a smooth $1/\sqrt{\en}$ thermal region, several wide low-energy resonances, a densely populated resolved resonance region and, finally, the unresolved resonance region. Therefore, this particular cross section will allow us to observe the response of the unfolding procedure to all the numerical conditions that may be expected from a typical neutron reaction dataset. In addition, Fig.~\ref{fig2} also shows (for visual purposes only) the portion of the folded spectrum inside the dense resolved resonance region, in order to emphasize the extreme smoothing effect that the resolution function commonly exhibits within this energy range.

\begin{figure*}[h!]
\includegraphics[width=0.5\linewidth]{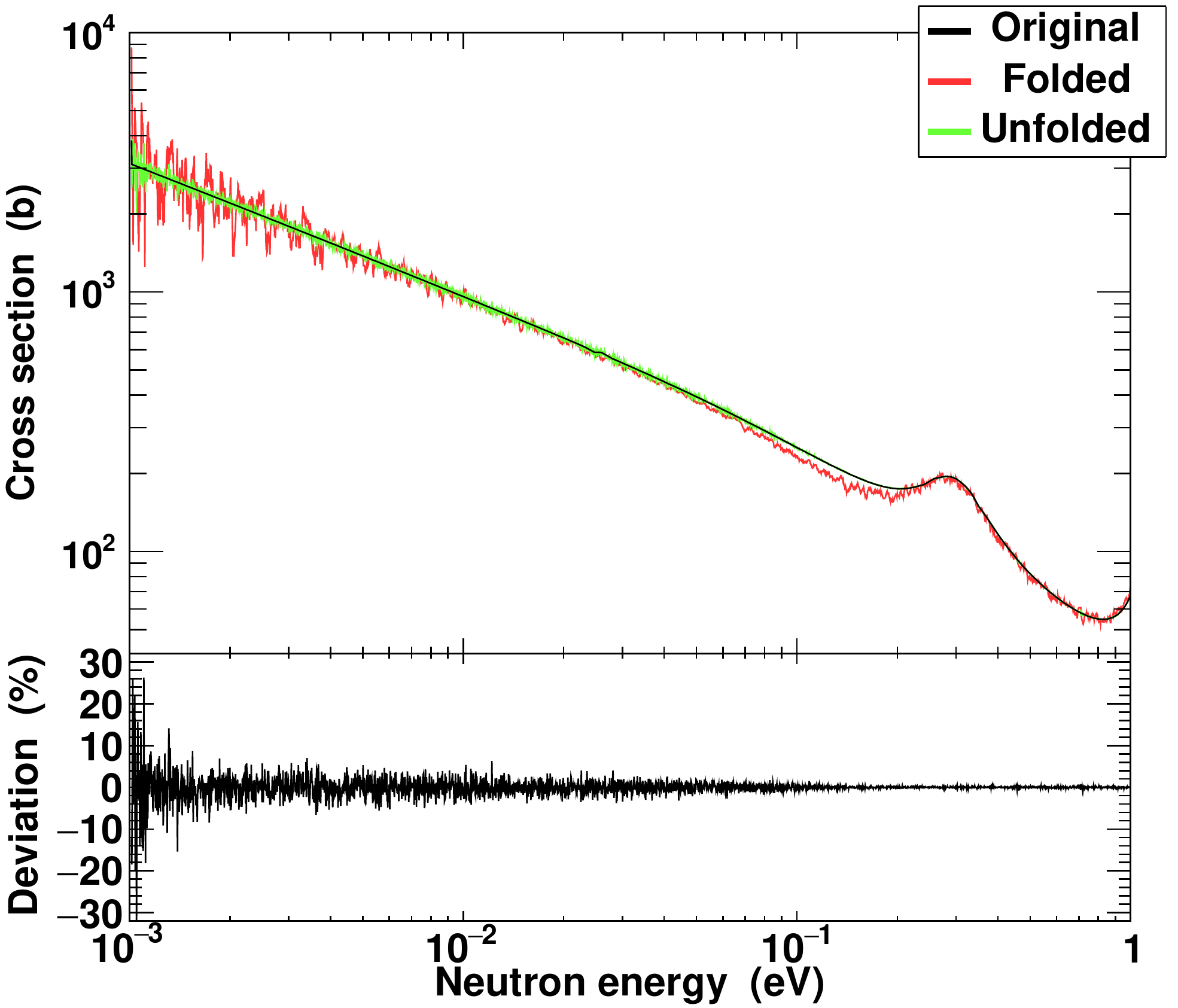}\includegraphics[width=0.5\linewidth]{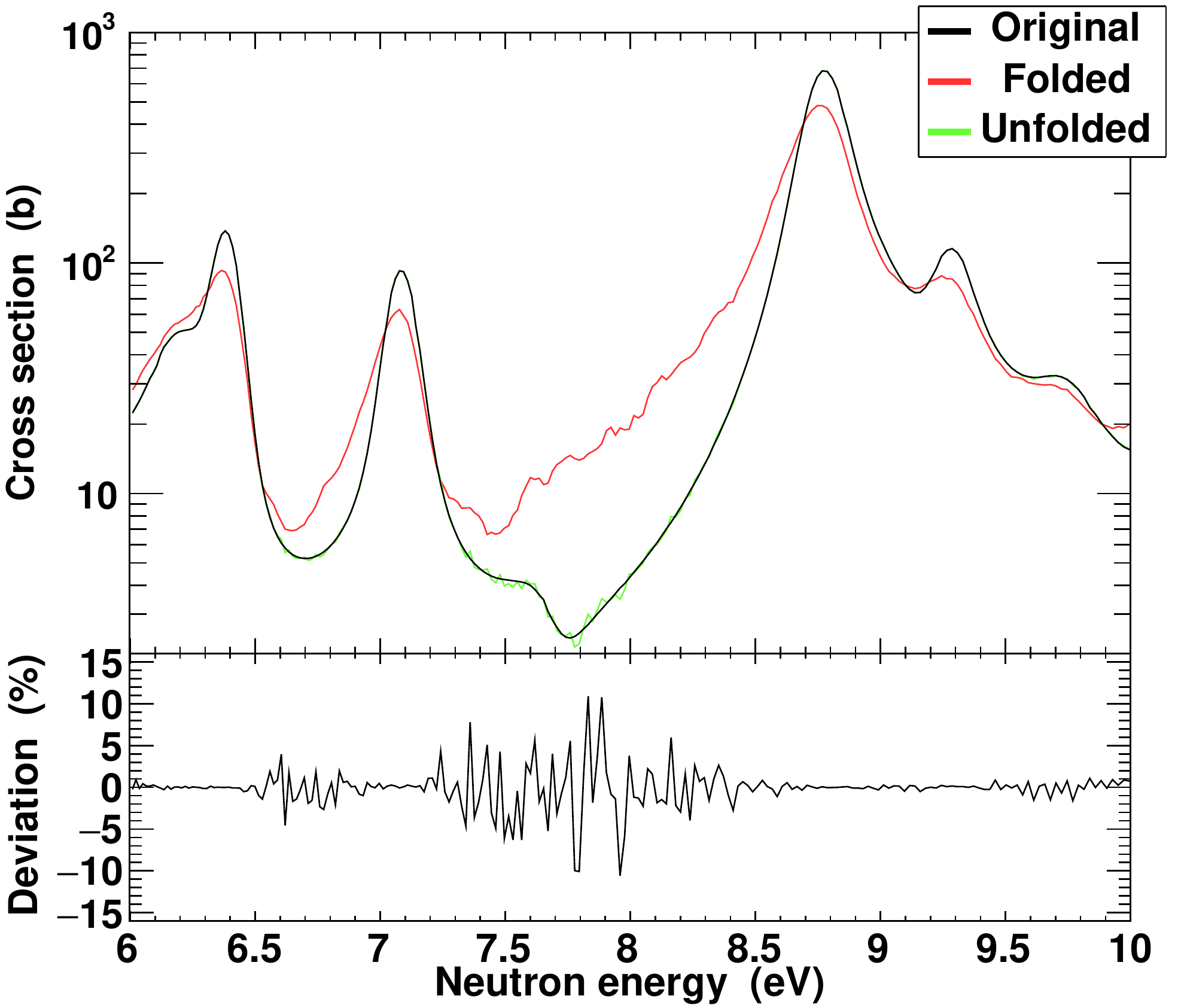}
\includegraphics[width=0.5\linewidth]{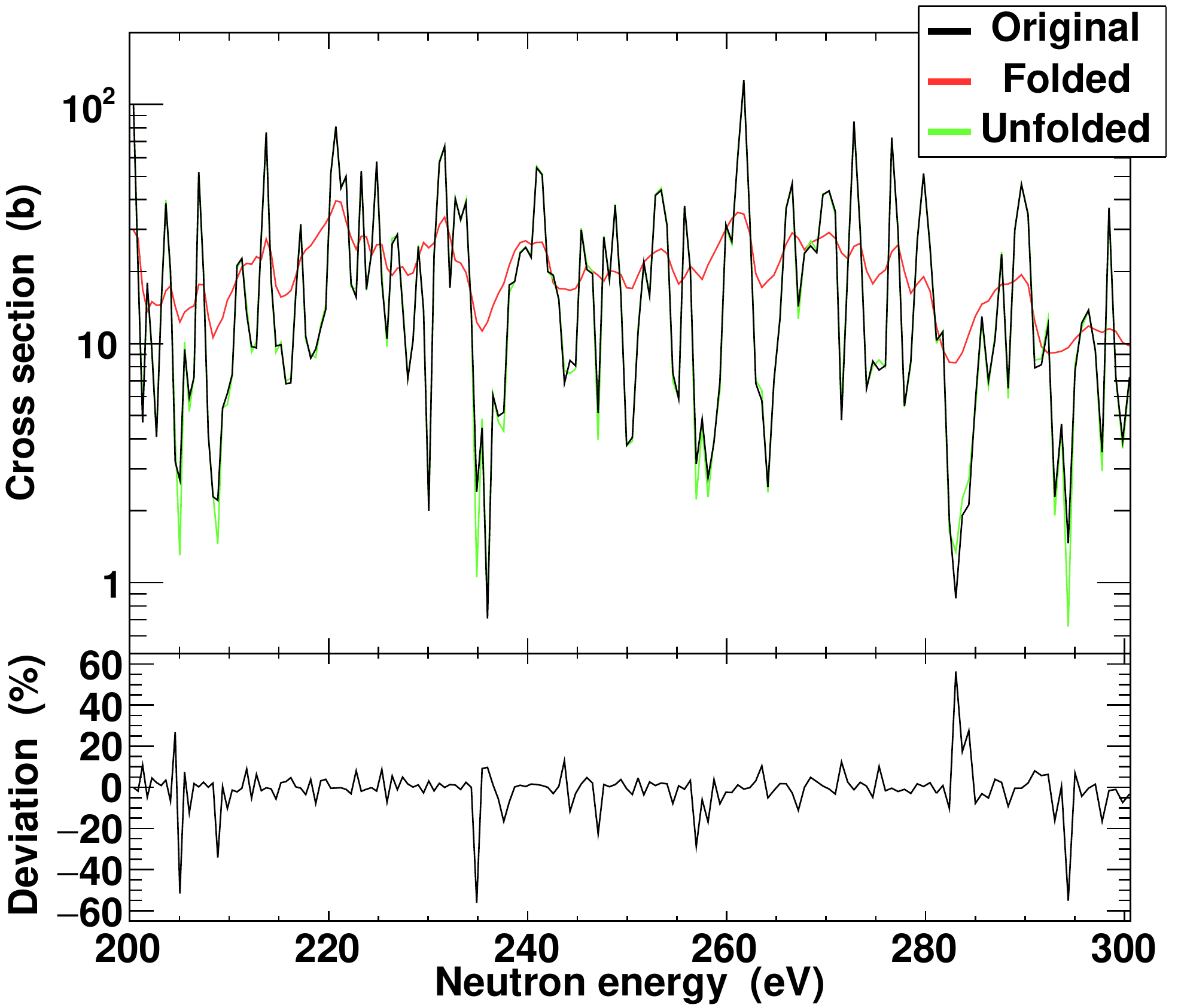}\includegraphics[width=0.5\linewidth]{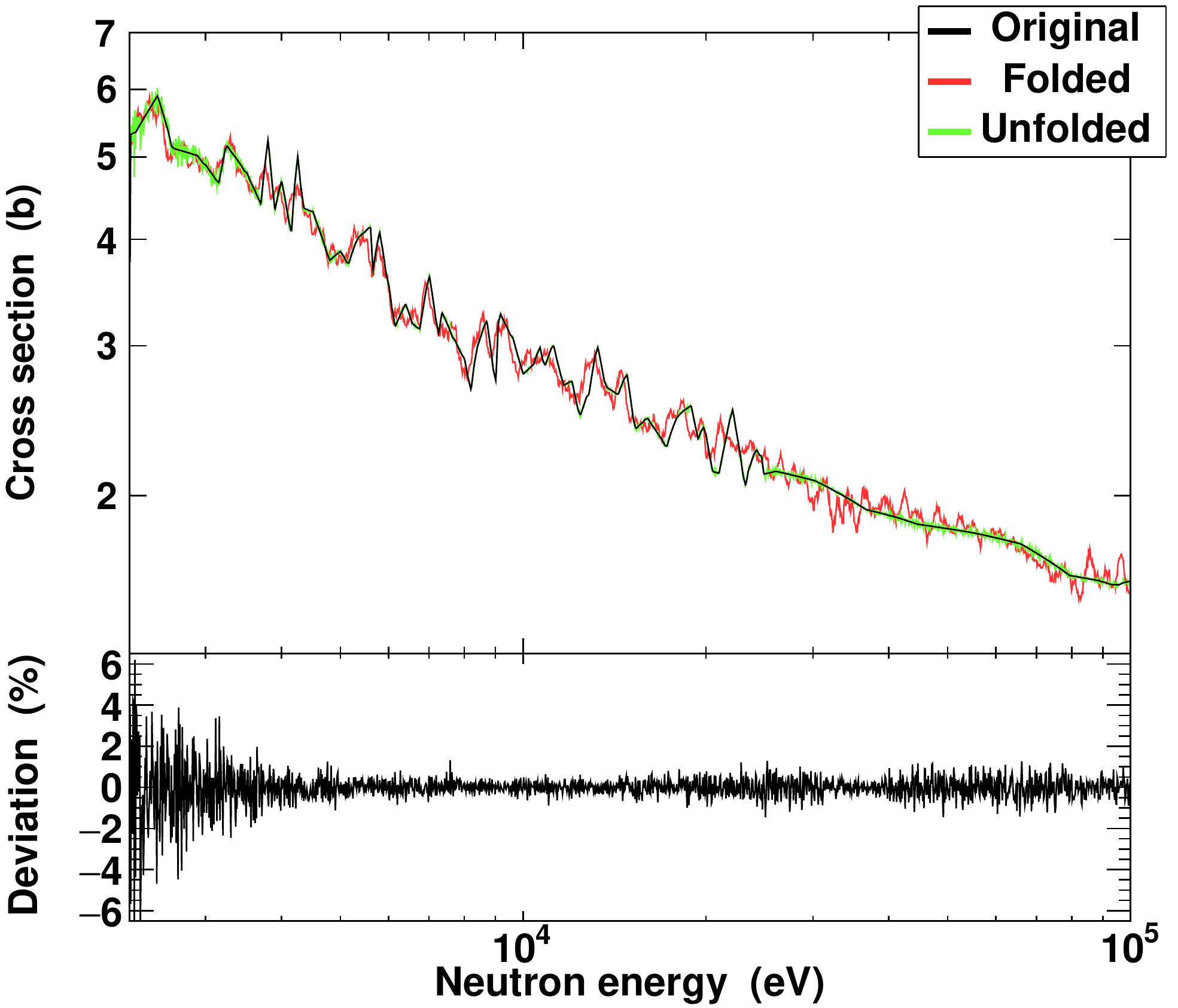}
\caption{Results of the resolution function unfolding procedure applied to the $^{235}$U($n$,f) cross section, within different representative energy regions. They are compared both to the spectrum folded by the resolution function for the EAR2 of the n\_TOF facility, and the true solution. Isolated subplots show the relative fluctuations of the unfolded spectrum around the true solution.}
\label{fig3}
\end{figure*}

We have applied the unfolding procedure to the spectra of maximum size of $10^5$ bins ($10^4$ bins per decade within the energy range from 1~meV to 10~MeV), which meant using the probability matrix $\pmx$ of size $10^5\times10^5$. Figure~\ref{fig3} shows the results for the $^{235}$U($n$,f) spectrum of this size. In order to reduce the visible dataset, in displaying the results we have kept only one in every ten points. As opposed to rebinning the dataset, which leads to a decrease in the relative fluctuations in the unfolded solution, rejecting not too large a portion of the points faithfully preserves the magnitude of these fluctuations (inasmuch as the remaining sample of points is representative of the entire population). The four plots from Fig.~\ref{fig3} focus on the smooth $1/\sqrt{E}$ range, wide low-energy resonances, densely populated resolved resonance region and the unresolved resonance region. Somewhat surprisingly, the unfolding procedure seems to suffer most in the smoothest parts of the spectrum, yielding the greatest fluctuations around the true solution (shown separately by the respective bottom panels). However, taking into account the size of the system, we consider the quality of the unfolded solution to be commendable. It is worth noting that, implemented in C++, the entire unfolding procedure -- which is by far dominated by the Cholesky decomposition -- takes 15~minutes for the system of size $10^5\times10^5$, on a single core of an Intel Core i5-6500 3.2GHz processor. On the other hand, the system of size $(5\times10^4)\times(5\times10^4)$ requires less than 2~minutes; the system of $(2\times10^4)\times(2\times10^4)$ takes 5~seconds, while the system of $10^4\times10^4$ is handled under a second.

\subsection{Iterative methods}
\label{iterative}

As an alternative to the direct methods of solving the linear system from Eq.~(\ref{new_master}) -- or out of necessity if none of them could be made to work -- one could always consider some of the iterative procedures. First among those is the \textit{conjugate gradient method}, requiring that the matrix be symmetric and positive definite \cite{matrix,numc}, which is a condition met by $\ppmx$. If (and only if) successful, for the system of size $n_\en\times n_\en$ the method is supposed to converge to an exact solution in at most $n_\en$ iterations. However, under the finite precision arithmetics the success of the method may strongly depend on the initial guess for the solution. In practical situations, this guess can hardly be anything other than $\vec{\mathcal{N}}_\x$ from the left side of Eq.~(\ref{new_master}). Unfortunately, we have observed that in our case the method starts diverging severely from the expected solution already for the system size of approximately $350\times350$. As such, we do not recommend it for solving this particular problem.

The next method worth considering is the \textit{successive over-relaxation} \cite{matrix,numc}. For the system $\ppmx \vec{x}=\vec{y}$ under consideration, its $k$-th iterative step may be expressed in a closed-form as:
\begin{linenomath}\begin{equation}
x_i^{(k)}=(1-\omega)x_i^{(k-1)}+\frac{\omega}{\ppmx_{ii}}\left(y_i-\sum_{j<i}\ppmx_{ij}x_j^{(k)}-\sum_{j>i}\ppmx_{ij}x_j^{(k-1)}\right)
\end{equation}\end{linenomath}
with $\omega$ as the relaxation parameter such that $0<\omega<2$. For a particular value of $\omega=1$ the method reduces to the so called \textit{Gauss-Seidel method}. The benefit of values different than unity may be the increased rate of convergence. However, the optimal value for $\omega$ strongly depends on the particular system being solved. We have found that the over-relaxation method works well in solving Eq.~(\ref{new_master}) even for the systems of the largest considered size ($10^5\times10^5$), thus it may be recommended for its success in unfolding the resolution function.

In practical applications no \textit{a priori} knowledge of the expected solution is available. For this reason, at every point we define a convergence criterion as the relative change $r_i$ between successive iterations:
\begin{linenomath}\begin{equation}
\label{crit}
r_i\equiv 2\left|\frac{x_i^{k}-x_i^{(k-1)}}{x_i^{k}+x_i^{(k-1)}}\right|
\end{equation}\end{linenomath}
and follow their evolution. Whenever the denominator happens to be 0, the corresponding $r_i$ are simply ignored. Requiring that all $r_i$ drop below a preset limit may be prohibitive, as there are always some sporadic far-away values to be expected. Alternatively, one could always require that some preset portion, e.g. 90\% of their distribution drops below a certain threshold. On the other hand, for a basically exponential-like shape of the distribution with a longer tail, we found that a good estimator of the relevant distribution range is $7\mu\sigma$, with $\mu$ as the mean value of $r_i$ values (excluding ignored ones) and $\sigma$ as their root mean square.

Despite the general success of the successive over-relaxation in unfolding the resolution function, we have found its rate of convergence to be rather slow. A carefully selected value of $\omega$ may, indeed, speed up the convergence rate. However, there is no unique value to be recommended, as it depends even on the number of bins in the analyzed spectra, i.e. on the size of the constructed $\ppmx$ matrix. Furthermore, we have observed the following: after reaching the same value of the convergence criterion using different values of $\omega$, the final iterative results are sometimes at visibly different levels of agreement with the expected solution. This difference is manifested through the increased or reduced residual fluctuations around the expected solution, but the quality of the particular solution does not seem to bear any correlation with the optimality of $\omega$ (in the sense of the convergence rate).

One can, of course, use the solution obtained by the Cholesky decomposition from Section~\ref{cholesky} as a starting point for any of the iterative procedures. If this option is available, i.e. if $\ppmx$ was, indeed, successfully decomposed so as to at least produce a numerically stable approximation to the solution of Eq.~(\ref{new_master}), then another type of an iterative improvement also becomes available. For a system $\ppmx \vec{x}=\vec{y}$ this improvement consists of a repeated application of the following procedure \cite{matrix,numc}:
\begin{linenomath}\begin{equation}
\label{iter}
\vec{x}_\mathrm{new}=\vec{x}_\mathrm{old}-\ppmx^{-1}\left(\ppmx\vec{x}_\mathrm{old}-\vec{y}\right)
\end{equation}\end{linenomath}
as many times as necessary, but rarely more than once. Equation~(\ref{iter}) is not to be interpreted as an algebraic identity yielding \mbox{$\vec{x}_\mathrm{new}=\ppmx^{-1}\vec{y}$} at every step, but as a literal instruction for the numerical evaluation of the right-hand side. In that, the term \mbox{$\delta\vec{x}=\ppmx^{-1}\left(\ppmx\vec{x}_\mathrm{old}-\vec{y}\right)$} should be calculated in the same way as the initial solution, i.e. by the forward and backward substitutions, taking advantage of already completed decomposition of $\ppmx$ in order to extract the solution of the system \mbox{$\ppmx\delta\vec{x}=\ppmx\vec{x}_\mathrm{old}-\vec{y}$} without explicitly constructing the inverse $\ppmx^{-1}$. Having said that, we have not found any significant improvement to the initial solution obtained directly from the Cholesky decomposition, using any combination of the iterative procedures. Therefore, we can wholeheartedly recommend the standalone application of the procedure from Section~\ref{cholesky} as selfsufficient.

\section{Uncertainty propagation}
\label{uncertainty}

The propagation of uncertainties is an important issue to be addressed in any kind of direct unfolding procedures. From the formal solution $\vec{N}_\en=\pmx^{-1}\:\vec{N}_\x$ to Eq.~(\ref{master}), it directly follows that the uncertainty $\sigma_i^{(\en)}\equiv[\vec{\sigma}^{(\en)}]_i$ of the $i$-th component of the reconstructed solution equals:
\begin{linenomath}\begin{equation}
\label{uncer}
\sigma_i^{(\en)}=\sqrt{\sum_j\left(\pmx_{ij}^{-1}\sigma_j^{(\x)}\right)^2}
\end{equation}\end{linenomath}
with $\sigma_j^{(\x)}\equiv[\vec{\sigma}^{(\x)}]_j$ as the uncertainty of the $j$-th component in $\vec{N}_\x$, together with the following notation: $\pmx_{ij}^{-1}\equiv[\pmx^{-1}]_{ij}$. The difficulty with evaluating Eq.~(\ref{uncer}) lies in the numerical efficiency because $\vec{\sigma}^{(\en)}$ \textit{can not} be expressed solely as a function of $\pmx^{-1}\vec{\sigma}^{(\x)}$, which we \textit{are} able to calculate efficiently by taking advantage of the Cholesky decomposition. In order to avoid an explicit calculation of $\pmx^{-1}$ -- just as we were striving to do up to this point -- we use Eq.~(\ref{uncer}) only as a starting point for developing the procedure that \textit{can} take advantage of the already performed decomposition. We first recognize that $\pmx^{-1}=\ppmx^{-1}\pmx^\top$, which leads to:
\begin{linenomath}\begin{equation}
\sigma_i^{(\en)}=\sqrt{\sum_j\left(\sum_k \ppmx_{ik}^{-1}\pmx_{kj}^\top \sigma_j^{(\x)}\right)^2}
\end{equation}\end{linenomath}
The benefit of defining the sequence of vectors $\vec{\chi}^{(j)}$:
\begin{linenomath}\begin{equation}
[\vec{\chi}^{(j)}]_k\equiv \pmx_{kj}^\top \sigma_j^{(\x)}
\end{equation}\end{linenomath}
soon becomes evident, as this formal manipulation allows us to write:
\begin{linenomath}\begin{equation}
\label{sigma}
\sigma_i^{(\en)}=\sqrt{\sum_j\left([\ppmx^{-1}\vec{\chi}^{(j)}]_i\right)^2}=\sqrt{\sum_j\left([\vec{\xi}^{(j)}]_i\right)^2}
\end{equation}\end{linenomath}
where we have immediately introduced an additional sequence of vectors $\vec{\xi}^{(j)}$, each of them to be found as a solution to the equation:
\begin{linenomath}\begin{equation}
\label{jots}
\ppmx\vec{\xi}^{(j)}=\vec{\chi}^{(j)}
\end{equation}\end{linenomath}
By again taking advantage of the already performed Cholesky decomposition of $\ppmx$, all $\vec{\xi}^{(j)}$ may be found just by using forward and backward substitutions, without ever explicitly constructing $\ppmx^{-1}$. In practical implementations $\vec{\xi}^{(j)}$ should not be \textit{all} computed first and only then Eq.~(\ref{sigma}) be evaluated. Such procedure would require a memory of size $n_\x\times n_\en$ ($n_\x$ vectors, each with $n_\en$ components), equivalent to storing the entire $\pmx$ matrix, thus defeating one of the main benefits of introducing the storage scheme from \ref{storage}. Instead, each $\vec{\xi}^{(j)}$ should be immediately used for incrementing all $n_E$ sums (one for every $\sigma_i^{(\en)}$), allowing to store only one $\vec{\xi}^{(j)}$ at a time.

Unfortunately, this exact procedure is of \mbox{$\mathcal{O}(n_\x\times n_\en\times B_{\ppmx})$} computational complexity -- $\mathcal{O}(n_\en\times B_{\ppmx})$ being the cost of solving each of the $n_\x$ systems from Eq.~(\ref{jots}), with $B_{\ppmx}$ as the bandwidth of $\ppmx$ -- and rapidly becomes prohibitive even for systems of modest size. Luckily, the exact procedure may be easily and reliably replaced by a rudimentary numerical simulation. Assuming that the original uncertainties are purely due to statistical fluctuations in the number of detected counts, a formally correct procedure would be to randomly generate across the whole spectrum a preselected number of $K$ sets of deviations $\vec{\Delta}_\x^{(k)}$ from the measured number of counts (\mbox{$k\in[1,K]$}), according to a Poisson distribution:
\begin{linenomath}\begin{equation}
[\vec{\Delta}_\x^{(k)}]_i=\mathrm{Poisson}\left(\mu=[\vec{N}_\x]_i\right)-[\vec{N}_\x]_i
\end{equation}\end{linenomath}
However, sampling the Gaussian distribution may be simpler and also works perfectly well:
\begin{linenomath}\begin{equation}
[\vec{\Delta}_\x^{(k)}]_i=\mathrm{Gauss}\left(\mu=0;\sigma^2=[\vec{N}_\x]_i\right)
\end{equation}\end{linenomath}
For each $\mbox{$k\in[1,K]$}$ a corresponding system:
\begin{linenomath}\begin{equation}
\ppmx\vec{\Delta}_\en^{(k)}=\pmx^\top\vec{\Delta}_\x^{(k)}
\end{equation}\end{linenomath}
is to be solved for $\vec{\Delta}_\en^{(k)}$ and the final set of uncertainties across the unfolded spectrum is to be calculated as:
\begin{linenomath}\begin{equation}
\label{sim_sig}
\sigma_i^{(\en)}=\sqrt{\frac{1}{K}\sum_{k=1}^{K}\left([\vec{\Delta}_\en^{(k)}]_i\right)^2}
\end{equation}\end{linenomath}
The computational complexity of this procedure is \mbox{$\mathcal{O}(K\times n_\en\times B_{\ppmx})$}, a significant improvement when the size $n_\x$ of the initial dataset is large. Though one should not settle for less than $K=100$ iterations, this value already yields a quite satisfactory level of precision in estimating the uncertainties that would otherwise be obtained by the exact procedure from Eq.~(\ref{sigma}).

In order to precisely evaluate severity by which the unfolding procedure affects the uncertainties, they need to be compared to the original uncertainties before unfolding. Here we need to take into account that the unfolded spectrum is always a function of the (true kinetic) neutron energy $\en$. Therefore, a direct comparison between the uncertainties before and after unfolding can only be performed if the original spectrum has also been constructed as a function of neutron energy (in this case the reconstructed energy $\etof$) and with the identical binning as the final unfolded spectrum $\vec{\sigma}^{(\en)}$, regardless of the kinematic parameter $\x$ and of the binning otherwise used for displaying the data before unfolding. Using the notation $\vec{\sigma}^{(\etof=\en)}$ for thus constructed set of uncertainties, the relative amplification $\rho_i$ of the uncertainties across the spectrum is:
\begin{linenomath}\begin{equation}
\label{ampl}
\rho_i\equiv\frac{\sigma_i^{(\en)}}{\sigma_i^{(\etof=\en)}}
\end{equation}\end{linenomath}
For a given binning, and as long as the original uncertainties are purely statistical (due to the fluctuations in the number of detected counts), these bin-wise amplification factors are insensitive to the total integrated number of counts, making them robust estimators of the effect of the unfolding procedure.

\begin{figure}[t!]
\includegraphics[width=1.0\linewidth]{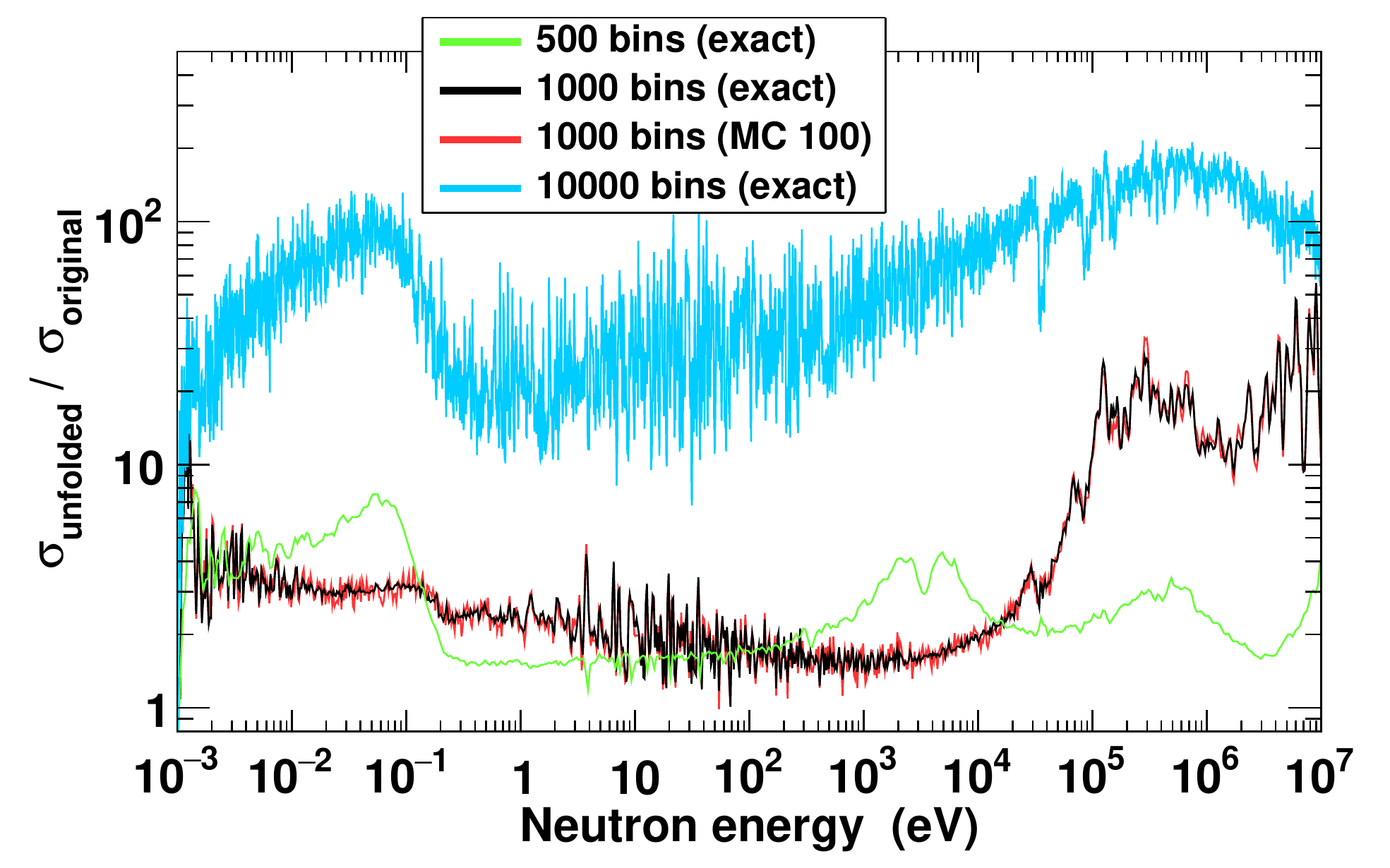}
\vspace*{-3mm}
\caption{Noise amplification induced by the direct unfolding procedure applied to the $^{235}$U($n$,f) cross section. For the spectrum of 1000 bins in total, simulated (MC: Monte Carlo) results obtained with 100 iterations are shown alongside the exact uncertainties.}
\label{fig4}
\end{figure}

Figure~\ref{fig4} shows the uncertainty amplification for the $^{235}$U($n$,f) cross section from Figs.~\ref{fig2}~and~\ref{fig3}, for three different binning densities: 50, 100 and 1000 bins per decade, i.e. 500, 1000 and 10000 bins in total. For the total of 1000 bins, the simulated results are also shown, obtained using $K=100$ iterations in the context of Eq.~(\ref{sim_sig}). Unfortunately, this increase in the uncertainty is not caused by the numerical stability, which could be improved by a more sophisticated algorithm. Rather, it is the general and inherent feature of the direct unfolding methods, i.e. of the direct inversion procedures. Therefore, while the method may work impressively for the spectrum entirely devoid of noise -- as demonstrated by Fig.~\ref{fig3} -- it can be meaningfully applied to the real-world data only if the noise is still at or below some acceptable level, even after amplification.

\section{Conclusions}
\label{conclusion}

We have explored a direct method of unfolding the resolution function from measurements of neutron induced reactions. The basic principle behind the method is straightforward and consists of solving a large linear set of equations, akin to a direct inversion of the resolution function matrix (RFM). The main concerns of the method are the numerical stability and the computational efficiency, since the systems of interest may become quite large, bringing the RFM closer and closer to being ill-conditioned. For large systems the computational efficiency requirement disqualifies some of the methods that would, in principle, be successful at solving the numerical stability issue, such as the Singular Value Decomposition. Fortunately, one may take advantage of the narrow banded structure of the RFM and combine the specialized matrix storage scheme, presented in this paper, with the Cholesky decomposition of the modified RFM in order to arrive at an efficient algorithm that was successfully applied to a system of size $10^5\times10^5$. However, a small modification of the large ill-conditioned RFM must be carried out in advance of the Cholesky decomposition, if the procedure is even to succeed, let alone be numerically stable during the inversion process. The modification consists of reinforcing the positive definiteness of the symmetric matrix $\ppmx$, introduced in Eq.~(\ref{new_master}). There is more than a single unique choice for doing this, and we have opted for a simple amplification of diagonal elements by a multiplicative factor $1+\epsilon$, using an optimal value of $\epsilon=10^{-4}$.

We have also explored the amenability of the problem to the iterative, rather than direct procedures. While the conjugate gradient method fails to properly reconstruct the solution already in the case of more than 350 equations, the successive overrelaxation method successfully handles the task, although at the price of a somewhat slow convergence. One could always use the solution obtained by the Cholesky decomposition as a starting point for any of the iterative methods. However, we have found this entirely unnecessary, due to the quality of the Cholesky solution already at the level where little could be gained from an iterative improvement.

A critical issue to be addressed regarding the direct unfolding methods is the propagation of uncertainties into the final solution. Unfortunately, the high sensitivity to the noise in the measured data is an inherent feature of such methods, as the noise is heavily amplified during the inversion procedure. Therefore, the level of the noise ultimately dictates if the procedure can be meaningfully applied. However, if the method \textit{is} to be applied, the expected increase in the noise can be exactly evaluated (given the known or expected distribution of uncertainties across the measured spectrum), thus allowing to determine in advance the acceptable level of the noise in the measured data.

{\color{white}.}

\textbf{ Acknowledgements}\\

This work was supported by the Croatian Science Foundation under Project No. 1680 and by the EC under Project FP7-PEOPLE-334315 ("NeutAndalus").

\appendix

\section{Matrix manipulations}
\label{matrix}

\subsection{Matrix storage}
\label{storage}

We will describe the storage scheme for banded matrices, which we apply to all matrices referred to in this work. Let $\mmx$ be the banded matrix of size $\mathcal{I}\times\mathcal{J}$. We will only store the consecutive sequence of elements from each column, outside of which all elements are 0. To this purpose we arrange the matrix content column-by-column into a one-dimensional array $M$, followed by the supporting arrays $i_\mn$ and $i_\mx$ containing the indices of the first and the last stored term from a given ($j$-th) column. In this sense $\mmx_{i_\mn[j],j}$ and $\mmx_{i_\mx[j],j}$ are the first and the last nonzero elements from the $j$-th column. While only these three arrays are necessary for a unique and unambiguous storage, it is extremely convenient to also maintain the arrays $j_\mn$ and $j_\mx$, containing the indices of the first and the last nonzero element from a given ($i$-th) row. The main reason is that this practice allows for a significant speedup in the multiplication of banded matrices: 
\begin{linenomath}\begin{equation}
(\amx\bmx)_{ij}=\sum_{k=\mx\big\{j_\mn^{\,(\amx)}[i],\:i_\mn^{\,(\bmx)}[j]\big\}}^{\mn\big\{j_\mx^{\,(\amx)}[i],\:i_\mx^{\,(\bmx)}[j]\big\}} \amx_{ik}\bmx_{kj}
\end{equation}\end{linenomath}
thus avoiding to loop over the off-band portions of either $\amx$ or $\bmx$, where the product $\amx_{ik}\bmx_{kj}$ is always~0.

A quick access to an arbitrary matrix element $\mmx_{ij}$ is facilitated by maintaining an additional array $I$, containing the indices of the positions in the array $M$ itself, where the content of a new column starts. In this case the efficient access is achieved as:
\begin{linenomath}\begin{equation}
\mmx_{ij}=\left\{\begin{array}{cc}M\left[I[j]+i-i_\mn[j]\right]& \;\mathrm{if}\; i_\mn[j]\le i \le i_\mx[j]\\0&\mathrm{otherwise}\end{array}\right.
\end{equation}\end{linenomath}
Evidently, one may store the matrix row-by-row, in which case an array $I$ needs to be replaced by a completely analogous array $J$, containing the indices of the positions within $M$ where the content of a new row starts. Hence, the access takes the form:
\begin{linenomath}\begin{equation}
\mmx_{ij}=\left\{\begin{array}{cc}M\left[J[i]+j-j_\mn[i]\right]& \;\mathrm{if}\; j_\mn[i]\le j \le j_\mx[i]\\0&\mathrm{otherwise}\end{array}\right.
\end{equation}\end{linenomath}
However, the column-by-column storage lends itself more naturally to the procedure ahead. It is worth noting that both pairs of index arrays -- $i_\mn$ with $i_\mx$ and $j_\mn$ with $j_\mx$ -- may be simultaneously updated while the matrix $\mmx$ is being constructed, regardless of the selected storage scheme. This fact may lead to the improvement in the computational efficiency when the construction of the next matrix element depends on the elements constructed up to that point, as during the various matrix decomposition operations.

In this work we adopt the convention of zero-offsetting all the array indices (implying that the first matrix term is $\mmx_{00}\leftrightarrow M[0]$). Recalling that the matrix $\mmx$ is of the size $\mathcal{I}\times\mathcal{J}$, we note the following: while the arrays $i_\mn$ and $i_\mx$ contain $\mathcal{J}$ elements each, their range of values spans between 0 and $\mathcal{I}-1$. Similarly, while the arrays $j_\mn$ and $j_\mx$ contain $\mathcal{I}$ elements, their values are bounded between 0 and $\mathcal{J}-1$. In the most general case, the only concern to be addressed is the arbitrary handling of the empty, i.e. all-zero rows or columns.

The very definition of the arrays $I$ and $J$:
\begin{linenomath}\begin{align}
\begin{split}
I[j]&=I[j-1]+\Big(i_\mx[j-1]-i_\mn[j-1]+1\Big)\\
J[i]&=J[i-1]+\Big(j_\mx[i-1]-j_\mn[i-1]+1\Big) 
\end{split}
\end{align}\end{linenomath}
with $I[0]=J[0]=0$, is also computationally the most efficient procedure. Due to the adopted zero-offset convention, the final terms in these arrays are $I[\mathcal{J}-1]$ and $J[\mathcal{I}-1]$. However, in the sense of the previous definitions, the terms $I[\mathcal{J}]$ and $J[\mathcal{I}]$:
\begin{linenomath}\begin{align}
\begin{split}
I[\mathcal{J}]&=\sum_{k=0}^{\mathcal{J}-1}\Big(i_\mx[k]-i_\mn[k]+1\Big)\\
J[\mathcal{I}]&=\sum_{k=0}^{\mathcal{I}-1}\Big(j_\mx[k]-j_\mn[k]+1\Big)
\end{split}
\end{align}\end{linenomath}
determine the total number of the elements stored in the array $M$, not all of which are necessarily non-zeros. It should be noted that in general, between the two storage schemes -- either by the columns or the rows -- the number of stored elements is not the same: $I[\mathcal{J}]\neq J[\mathcal{I}]$, which is easily proven by considering the minimal matrix example \mbox{$\mmx=[\emptyset\;0\;\emptyset]$}. If no dummy elements are used to fill-in all-zero columns or rows within the array $M$, here it holds: $I[\mathcal{J}=3]=2$ and $J[\mathcal{I}=1]=3$.

\subsection{Matrix preparation}
\label{preparation}

Before finding the solution to Eq.~(\ref{master}), several technical procedures need to be performed:
\begin{itemize}
\itemsep-0.5ex
\item[(1)] construction of an appropriate matrix $\pmx$
\item[(2)] reduction of $\pmx$
\item[(3)] localization of the relevant invertible subportion of $\pmx$
\item[(4)] normalization of $\pmx$
\end{itemize}
We address these procedures one by one.\\

\textbf{Constructing $\pmx$.} The only reliable way to determine the resolution function over the wide range of energies covered at n\_TOF is by dedicated simulations of the neutron production and transport process. The initial form $\pmx_0$ of the probability matrix is then obtained by directly histogramming the simulated data. However, this procedure can not be relied on in case of an arbitrarily high binning, as the amount of statistics acquired from the simulations is limited. Therefore, for each specific binning the matrix $\pmx$ should be constructed by an appropriate interpolation, starting from an optimally binned $\pmx_0$. Regardless of the selection of the kinematic parameter $\x\in\{\tf,\etof,\lambda\}$ used for expressing the initial $\pmx_0$, the transition into the selected functional dependence of the final $\pmx$ may be efficiently performed in a course of a single interpolation process, without the need for constructing any intermediate forms of either $\pmx_0$ or $\pmx$.

\begin{figure}[b!]
\includegraphics[width=1.0\linewidth]{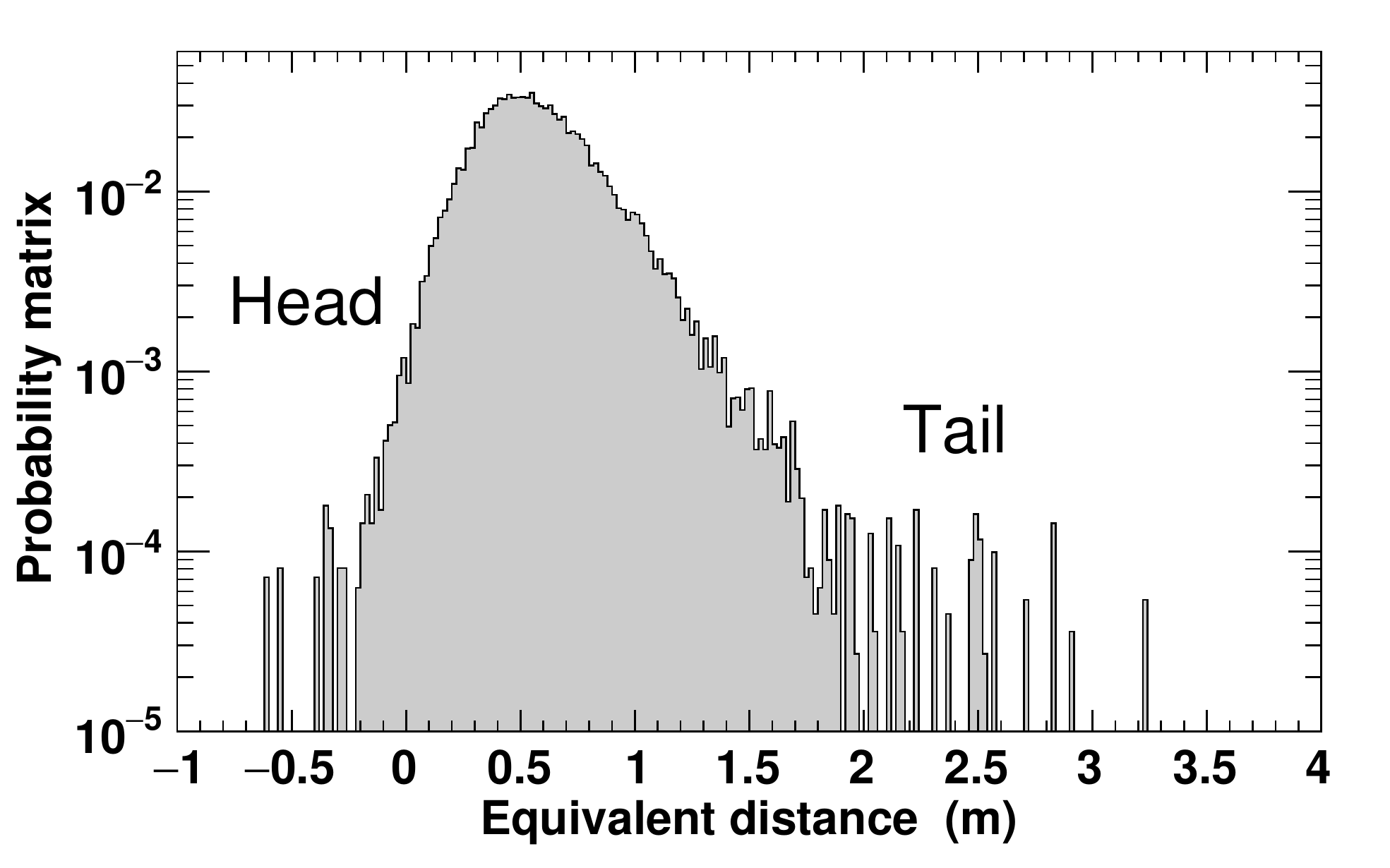}
\caption{Example of the slice through the probability matrix for the EAR2 of the n\_TOF facility, at the neutron kinetic energy of \mbox{$\en=2$~MeV}. The full probability matrix is shown by the top panel from Fig.~\ref{fig1}.}
\label{figx}
\end{figure}

\textbf{Reducing $\pmx$.} When the data from simulations are histogrammed, there will always be few sporadic counts at the tails of the distribution (in the direction of the kinematic parameter $\x$), which unnecessarily increase the column-wise bandwidth (the span between $i_\mn[j]$ and $i_\mx[j]$), but do not contribute in any significant way to the total resolution function. For this reason we recommend to reduce the stored content of $\pmx$ by cutting away the edges of the initially constructed matrix. This procedure is even more important if $\pmx$ was initially constructed not by directly histogramming the simulated data, but by evaluating the global analytical fit to the simulated data, such as the one proposed in Ref.~\cite{ntof} for modeling the n\_TOF resolution function. In this case the analytical form may be evaluated at every point throughout the entire $\pmx$ matrix, destroying its banded structure and defeating any benefit of the storage scheme from \ref{storage}. Therefore, we keep only the most relevant portion of the resolution function. Some consideration is required, though, as $\pmx$ is regularly asymmetric along the $\x$-direction. For $X=\tf$ the probability distribution has a long tail in the direction of longer flight times, while for $\x=\etof$ the tail extends towards the lower reconstructed energies. However, neither is the head of the distribution sharp, just more quickly decreasing than the tail. Therefore, the distribution needs to be cut at both sides, but in a way that the reduction at the tail is given precedence. For the clarity of terminology, Fig.~\ref{figx} illustrates the difference between the head and the tail of the distribution. Though the reduction is to be performed either in the time of flight or the reconstructed energy spectrum, Fig.~\ref{figx} shows the distribution as a function of the effective neutron flight path.

For the matrix reduction we employ the following simple algorithm. Let $\delta$ be the maximum allowed portion of probability to be discarded from each column of $\pmx$ (i.e. for each value of the true kinetic energy $\en$). For each column let: $\Sigma_\mathrm{total}=\sum_i \pmx_{ij}$ (at this point the columns of $\pmx$ have not yet necessarily been normalized to unity). Then for each column do the following:
\begin{itemize}
\itemsep-0.5ex
\item[(1)] keep discarding the elements from the head of the distribution as long as the sum $\Sigma_\mathrm{head}$ of the discarded elements is lower than the half of the assigned portion: \mbox{$\Sigma_\mathrm{head}<(\delta/2)\times\Sigma_\mathrm{tot}$}
\item[(2)] keep discarding the elements from the tail as long as the total sum of discarded elements does not exceed the assigned portion: \mbox{$\Sigma_\mathrm{head}+\Sigma_\mathrm{tail}\le\delta\times\Sigma_\mathrm{tot}$}
\item[(3)] continue discarding the elements from the head as long as the assigned portion is not exceeded: \mbox{$\Sigma_\mathrm{head}+\Sigma_\mathrm{tail}\le\delta\times\Sigma_\mathrm{tot}$}
\end{itemize}

At the end of the procedure it is possible that the greater contribution will be discarded from the head than from the tail of the distribution ($\Sigma_\mathrm{head}>\Sigma_\mathrm{tail}$). However, the tail is given a greater chance of dominating the total discarded content. For this work we have selected a value of $\delta=1\%$. Note that this procedure may be performed during the column-by-column construction of $\pmx$, meaning that one does not need to construct the entire matrix first and cut it afterwards, which enables the construction of matrices that might not fit the available memory resources before cutting. This possibility is greatly facilitated by the storage scheme from \ref{storage}.

\textbf{Localizing $\pmx$.} One needs to ensure that $\pmx$ does not span the range outside the available resolution function data. Even a single row or column in $\pmx$ completely composed of zeros implies that $\pmx$ is singular, hence uninvertible. In order to find the relevant invertible subportion of $\pmx$, we first determine the "center of mass" coordinates $\bar{\imath}$ and $\bar{\jmath}$:\footnote{If $\pmx$ were already normalized at this point, one might be tempted to reduce the denominator to $\sum_{i,j}\pmx_{ij}=n_\en$ and even to obtain \mbox{$\sum_{i,j} j\times\pmx_{i,j}=(n_\en-1)/2$}. Alas, this does not work precisely when it matters: when the column is completely filled with zeros $\sum_i\pmx_{ij}=1$ does not hold any more for a given~$j$.}
\begin{linenomath}\begin{equation}
(\bar{\imath},\bar{\jmath})=\frac{\sum_{i,j} (i,j)\times\pmx_{ij}}{\sum_{i,j}\pmx_{ij}}
\end{equation}\end{linenomath}
Starting from the position $(\bar{\imath},\bar{\jmath})$, we search for the nearest rows and columns containing all zeros. Supposing that such rows and/or columns exist, we reduce $\pmx$ in size so that all the content beyond these rows and columns is discarded. The procedure needs to be repeated until no such rows/columns exist any more, since reducing the matrix in such a way may introduce new all-zero rows/columns when there were none before. This happens when the only non-zero entries from these rows/columns have all been discarded by the last iteration of the matrix localization. It should be noted that this procedure may also be efficiently performed in-place just by shifting the matrix content, without constructing any intermediate containers. This is again greatly facilitated by the storage scheme from \ref{storage}.

\textbf{Normalizing $\pmx$.} Only at this point should the probability normalization ($\sum_i\pmx_{ij}=1$ for every $j$) be performed. Otherwise, the total sum of probabilities in each column would be reduced by the portion of the discarded content (contributed both by the rejection factor $\delta$ and the portions of $\pmx$ discarded during the localization process), thus implying the nonconservation of counts when $\pmx$ or, conversely, $\pmx^{-1}$ is applied.

\section{Multiple scattering effects}
\label{multiple}

In the presence of the pronounced multiple scattering effects the direct resolution function unfolding can not be meaningfully applied, due to their coupling not being multiplicatively separable. The inconvenience in obtaining the full and detailed parameterization of the multiple scattering effects stems from their dependence on every single sample -- its cross section, shape, size, mass and material homogeneity -- unlike the resolution function, which is uniquely determined by the neutron production facility.

In order to demonstrate this fact, let us consider the total number of detected counts $\D N_\en(\en)$ caused by the neutrons of energy $\en$. While they are expected to have been measured with the time of flight $\tf_\en=L/v_\en$, $v_\en$ being the neutron speed defined by Eq.~(\ref{lambda}), they are first delayed or advanced in time by $\tau_R$ due to the resolution function $R_\tf(\tf_\en+\tau_R,\en)$, only to be further offset by $\tau_M$ due to the multiple scattering effect $M_\tf(\tf_\en+\tau_M,\en)$. Therefore the total number of $\D^3N(\tf,\en)$ of counts detected with the time of flight $\tf$:
\begin{linenomath}\begin{equation}
\label{tof}
\tf=\tf_\en+\tau_R+\tau_M
\end{equation}\end{linenomath}
but offset specifically by $\tau_R$ and $\tau_M$ equals:
\begin{linenomath}\begin{equation}
\label{ap_d3}
\D^3N(\tf,\en)=R_\tf(\tf_\en+\tau_R,\en)\times M_\tf(\tf_\en+\tau_M,\en)\times\D\tau_R\D\tau_M\D N_\en(\en)
\end{equation}\end{linenomath}
Adopting the substitution:
\begin{linenomath}\begin{equation}
\left.\begin{array}{c}
T_R\equiv \tf_\en+\tau_R\\
T_M\equiv \tf_\en+\tau_M
\end{array}\right\} \quad\Rightarrow\quad \tf=T_R+T_M-\tf_\en
\end{equation}\end{linenomath}
and transitioning from $T_M$ to $\tf$, Eq.~(\ref{ap_d3}) becomes:
\begin{linenomath}\begin{equation}
\D^3N(\tf,\en)=R_\tf(T_R,\en)\times M_\tf(\tf+\tf_\en-T_R,\en)\times\D T_R\D \tf\:\D N_\en(\en)
\end{equation}\end{linenomath}
Finally, to remove the experimentally indistinguishable sensitivity to the pure resolution function effect, separate from the multiple scattering contribution to the measured time of flight $\tf$, an integration over $T_R$ needs to be performed:
\begin{linenomath}\begin{equation}
\label{d3}
\D^2N(\tf,\en)=\D \tf\:\D N_\en(\en)\int_{-\infty}^\infty R_\tf(T_R,\en)\times M_\tf(\tf+\tf_\en-T_R,\en)\times\D T_R
\end{equation}\end{linenomath}
from where it is evident that in place of the resolution function from Eq.~(\ref{eq_d2}), a joint effect $\q_\tf(\tf,E)$ remains:
\begin{linenomath}\begin{equation}
\q_\tf(\tf,E)\equiv\int_{-\infty}^\infty R_\tf(T_R,\en)\times M_\tf(\tf+\tf_\en-T_R,\en)\times\D T_R
\end{equation}\end{linenomath}
Since the form is multiplicatively inseparable ($\qmx\neq\rmx\mmx$ for corresponding matrices), the resolution function can not be directly separated from the measurements, without having a full parameterization of the multiple scattering effects at hand.

\end{document}